\documentclass[fleqn,10pt]{SelfArx} 

\usepackage[english]{babel} 
\usepackage[super,compress]{cite}
\usepackage[justification=centerlast]{caption}[2020/09/12]

\definecolor{color1}{RGB}{0,0,90} 
\definecolor{color2}{RGB}{0,20,20} 

\usepackage{hyperref} 

\hypersetup{
	hidelinks,
	colorlinks,
	breaklinks=true,
	urlcolor=color2,
	citecolor=color1,
	linkcolor=color1,
	bookmarksopen=false,
	pdftitle={Title},
	pdfauthor={Author},
}

\JournalInfo{Condensed Matter -- Materials Science} 
\Archive{arXiv.org} 

\PaperTitle{Covalent Organic Functionalization of Graphene Nanosheets and Reduced Graphene Oxide via 1,3-Dipolar Cycloaddition of Azomethine Ylide} 

\Authors{Luca Basta\textsuperscript{1}*, Aldo Moscardini\textsuperscript{1}, Filippo Fabbri\textsuperscript{1}, Luca Bellucci\textsuperscript{1}, Valentina Tozzini\textsuperscript{1}, Silvia Rubini\textsuperscript{2}, Andrea Griesi\textsuperscript{3,4}, Mauro Gemmi\textsuperscript{4}, Stefan Heun\textsuperscript{1}, Stefano Veronesi\textsuperscript{1}\ddag} 
\affiliation{\textsuperscript{1}\textit{NEST, Istituto Nanoscienze-CNR and Scuola Normale Superiore, Piazza S. Silvestro 12, 56127 Pisa, Italy}} 
\affiliation{\textsuperscript{2}\textit{Istituto Officina dei Materiali CNR, Laboratorio TASC, Area Science Park - S S 14, km 163.5, I-34012 Trieste, Italy}} 
\affiliation{\textsuperscript{3}\textit{Department of Chemistry, Life Sciences and Environmental Sustainability, University of Parma, Parco Area delle Scienze 17/A, 43124 Parma, Italy}} 
\affiliation{\textsuperscript{4}\textit{Center for Nanotechnology Innovation@NEST, Istituto Italiano di Tecnologia, Piazza S. Silvestro 12, 56127 Pisa, Italy}} 
\affiliation{*\textbf{Corresponding author}: E-mail: luca.basta1@sns.it} 
\affiliation{\ddag\textbf{Corresponding author}: Tel: +39 050 509882. E-mail: stefano.veronesi@nano.cnr.it} 

\newcommand{\keywordname}{Keywords} 

\Abstract{Organic functionalization of graphene is successfully performed \textit{via} 1,3-dipolar cycloaddition of azomethine ylide in the liquid phase. The comparison between 1-methyl-2-pyrrolidinone and N,N-dimethylformamide as dispersant solvents, and between sonication and homogenization as dispersion techniques, proves N,N-dimethylformamide and homogenization as the most effective choice. The functionalization of graphene nanosheets and reduced graphene oxide is confirmed using different techniques. Among them, energy-dispersive X-ray spectroscopy allows to map the pyrrolidine ring of the azomethine ylide on the surface of functionalized graphene, while micro-Raman spectroscopy detects new features arising from the functionalization, which are described in agreement with the power spectrum obtained from ab initio molecular dynamics simulation. Moreover, X-ray photoemission spectroscopy of functionalized graphene allows the quantitative elemental analysis and the estimation of the surface coverage, showing a higher degree of functionalization for reduced graphene oxide. This more reactive behavior originates from the localization of partial charges on its surface due to the presence of oxygen defects, as shown by the simulation of the electrostatic features. Functionalization of graphene using 1,3-dipolar cycloaddition is shown to be a significant step towards the controlled synthesis of graphene-based complex structures and devices at the nanoscale.}

\begin{document}

\maketitle 


\thispagestyle{empty} 


\section*{Introduction} 

Since Novoselov and Geim isolated graphene for the first time in 2004,\cite{Novoselov04} many research efforts have been directed to its study, and remarkable results have been achieved in recent years.\cite{Chang13,Ferrari15,Yu20progress,Samori20introduction} Graphene is the first discovered two-dimensional atomic crystal. It is composed of carbon atoms packed in a honeycomb lattice.\cite{Geim07} Because of its unique one atom-thick structure, it combines outstanding properties such as superior mechanical strength,\cite{Lee08} excellent flexibility,\cite{Koenig11,Zhang11} high transparency,\cite{Nair08} exceptionally high thermal and electrical conductivities as well as amazing electronic properties.\cite{Balandin08,Du08,Chen11,Seol10,Zhang05,Stander09,Bolotin09,Novoselov05,Berger06}

Notwithstanding graphene's great application potential,\cite{Yu17} its outstanding properties are also a limitation. Due to the absence of a band-gap, graphene's use as an active element in electronic devices and sensors encounters many limitations and, therefore, some sort of engineering of graphene's structure is required.\cite{Schwierz10} It has been suggested that a powerful method to overcome these impediments is provided by an efficient surface functionalization with suitable materials. Functionalization of graphene offers the possibility to finely tune the system's physical and chemical properties, resulting in a synergistic combination of the individual features of each component. However, while graphene's high specific surface area of $2630$~$\mathrm{m^2/g}$\cite{Ambrosi14} provides numerous possible binding sites, its chemical inertness makes it difficult to modify graphene's structure without disrupting it or introducing excessive disorder.\cite{Mohan18} Notably, graphene composites have been tailored to have desired solubility and stability,\cite{Mendez19,Zhang11novel} tunable optical, electric, thermal, and mechanical properties,\cite{Loh10,Chiou12,Kim13,Goods14,Uddin14} enhanced catalytic capability,\cite{Roy10,Bakandritsos19} and biological interactions.\cite{Roy10,Jiang11,Lu21,Catania21review} Moreover, where the 3-dimensionality of graphene-based materials becomes fundamental, like in gas and energy storage applications,\cite{Bellucci20silico,Tozzini13,Zhu14,Bonaccorso15} the modification of graphene's surface with heteroatoms or functional groups allows enhanced performance.\cite{Dai13,Mashoff13,Wei17,Takahashi16,Sun18,Bakandritsos18,Bellucci20engineering} Nonetheless, to finely control or intentionally design the binding sites of functionalizing molecules on graphene's surface while preserving the high quality of its unique structure remains an open challenge.

Generally, graphene functionalization can be achieved through either covalent bonds or non-covalent in\-ter\-actions.\cite{Liu12,Hirsch13} The covalent functionalization is usually realized via substitution of carbon atoms in the basal plane of graphene by heteroatoms (atom doping),\cite{Hu18,Boukhvalov12} reaction with residual functional groups on graphene,\cite{Melucci10} or modification of graphene's unsaturated structure.\cite{Feng13,Bueno17,Lu18,Park13,Criado15,Stergiou20} In contrast, the non-covalent chemical modification is often realized by intermolecular interactions such as van der Waals forces,\cite{Woszczyna14} electrostatic interactions,\cite{Laaksonen10} or $\pi-\pi$ stacking interaction.\cite{Bai09,Arranz20,Georgakilas16}

In particular, the covalent functionalization of graphene sheets using organic functional groups has been explored as a pivotal step towards the formation of graphene composites at the nanoscale. A commonly diffuse approach is the use of diazonium salts, often done through electrochemical or heating processes. Upon surface reduction of the diazonium compound, a highly reactive free radical is produced, which attacks the sp\textsuperscript{2} carbon atoms of the graphene lattice, forming a covalent bond.\cite{Lomeda08,Lim10,Huang13,Li19,Ambrosio20impact} While this mechanism leads to an abundant and quick functionalization, it is hard to control. Therefore, a more selective and controlled method has been shown as very promising. 1,3-dipolar cycloaddition (1,3-DC) of azomethine ylide has been investigated for the chemical modification of carbon nanotubes, fullerenes, and other carbon nanostructures,\cite{Georgakilas02,Tasis06,Kostarelos07,Maggini93,Tagmatarchis03,Cioffi06} thanks to the availability of a variety of organic derivatives by selecting the appropriate precursors, with interesting applications in different fields like biotechnology, sensors at the nanoscale, and solar energy conversion.\cite{Prato97,Tagmatarchis04,Giofre20eco} Recently, the graft of azomethine ylide on graphene via 1,3-DC has been reported after the successful production of graphene sheets directly from graphite dispersed in organic solvents.\cite{Georgakilas10} That was a pioneering work, but lacking a systematic and wide investigation on solvent effects, computational investigation, XPS analysis, or specific Raman signature of the functionalization. In a similar way, the selective binding between the amino groups involved in the 1,3-DC on dispersed graphene and gold nanorods has been obtained.\cite{Quintana10} However, a detailed description of graphene functionalized with azomethine ylide is still missing, as well as a deeper investigation on functionalization efficiency and signature.

While the techniques of production and chemical modification of graphene-like systems advance, the choice of the organic solvent suitable for the dispersion, together with the detailed understanding of the system, becomes fundamental in order to obtain an efficient functionalization. Here we present a detailed investigation on 1,3-DC of azomethine ylide on both graphene nanosheets (GNS) and reduced graphene oxide (rGO), including solvent effects. We compare sonication and homogenization as dispersion techniques, while 1-methyl-2-pyrrolidinone (NMP) and N,N-dimethylformamide (DMF) are used as dispersant solvents, for their excellent efficiency in dispersing graphene as well as facilitating the in situ production and the grafting of the ylide onto graphene. Energy-dispersive X-ray spectroscopy (EDX) and electron energy loss spectroscopy (EELS) confirm the organic functionalization of graphene and verify its homogeneity on the entire surface. While these techniques have already been applied to analogous systems, a detailed characterization of functionalized graphene with Raman spectroscopy is presented here for the first time. Raman spectroscopy, performed on both the pristine and functionalized samples, allowed the detection of the characteristic Raman signature of graphene together with new distinctive peaks from the ylide. Moreover, X-ray photoemission spectroscopy (XPS) is used to estimate the coverage and the efficiency of the functionalization process, assessing the elemental composition. An ab initio density-functional theory (DFT) molecular dynamics simulation illustrates how the presence of epoxy groups in the rGO surface induces a local inhomogeneity of the partial charges of gaphene's structure, promoting the 1,3-DC of azomethine ylide and showing how defects (like functional groups, edges, or vacancies) could be exploited to acquire control on the degree of the functionalization. Finally, the simulated power spectrum provides a precise idea of the Raman signature of functionalized graphene, in agreement with the experimental data.

\section{Results and Discussion}

\addcontentsline{toc}{section}{Results and Discussion} 

\subsection{Dispersion of GNS}
\addcontentsline{toc}{subsection}{Dispersion of GNS} 

Exfoliated GNS produced by wet-jet milling (as described elsewhere)\cite{Castillo18} were dispersed in NMP and DMF in order to obtain a stable dispersion ($\sim$ 0.2 mg mL\textsuperscript{-1}), as shown in Figure S1 in the Supporting Information. It is well known that these are well-suited organic solvents for the dispersion of graphene,\cite{Ciesielski14} since they minimize the interfacial tension between solvent and graphene. {Moreover, both NMP and DMF are commonly used in 1,3-dipolar cycloadditions for their ability to favor the reaction.\cite{Georgakilas10, Quintana10, Mali171, Chen19substituent} This is fundamental during the functionalization process, which is described in the following sections. In order to determine the best dispersant solvent and the most efficient dispersion technique, DLS measurements were performed on pristine GNS. Comparing the average lateral dimension of the dispersed nanosheets, we noticed a reduction in size from 850 nm to 600 nm (using sonication) or 400 nm (using homogenization) in NMP, in contrast with a decrease in size from 950 nm to 850 nm (using sonication) or 600 nm (using homogenization) in DMF (more details in the Supporting Information, Table S1). From these measurements, NMP occurs to be a slightly better dispersant, as a result of its optimal surface tension ($\gamma$ = 40 mJ m\textsuperscript{-2}), with respect to DMF ($\gamma$ = 37.1~mJ~m\textsuperscript{-2}).\cite{Hernandez08}
	
Sonication has been widely utilized to induce exfoliation of bulk materials through growth and collapse of micro-bubbles due to pressure fluctuations in liquids.\cite{Xia13} Recently, also homogenization, which consists in the shear mixing in suitable stabilizing liquids, has been demonstrated to achieve excellent exfoliation of graphite, even in a more scalable way.\cite{Paton14} Here, we observed that in both NMP and DMF, homogenization allows a faster dispersion of GNS. For example, sonication in NMP for 60 minutes allows a decrease of the average lateral dimension from 850 nm to 600~nm, while the same result was obtained after only 30 minutes of homogenization. Therefore, homogenization results to be the advisable dispersion procedure, and the subsequent experiments were performed after dispersion via homogenization.
	
\subsection{1,3-DC of GNS and rGO}
	
\addcontentsline{toc}{subsection}{1,3-DC of GNS and rGO} 
	
Functionalized GNS were prepared by adding N-methylglycine and 3,4-di\-hy\-droxy\-ben\-zal\-de\-hyde to the GNS dispersion, as schematically represented in Figure \ref{fig:DCscheme}. The 1,3-DC of GNS was performed both in NMP and in DMF, in order to allow a comparison between the two solvents. Notably, a higher degree of functionalization was achieved for GNS dispersed in DMF. Moreover, together with their dispersant efficiency, an important parameter to take into account is the boiling point of the solvents, 202~°C for NMP and 153 °C for DMF. A higher boiling point results in a more time-consuming procedure for removing the solvent under vacuum after the dropcasting of the functionalized graphene onto the substrates for characterization. Taking all this into consideration, rGO was dispersed only in DMF ($\sim$ 0.2 mg mL\textsuperscript{-1}) and only via homogenization, while the subsequential functionalization procedure remained the same.
	
\begin{figure}[h]
		\centering
		\includegraphics[width=\linewidth]{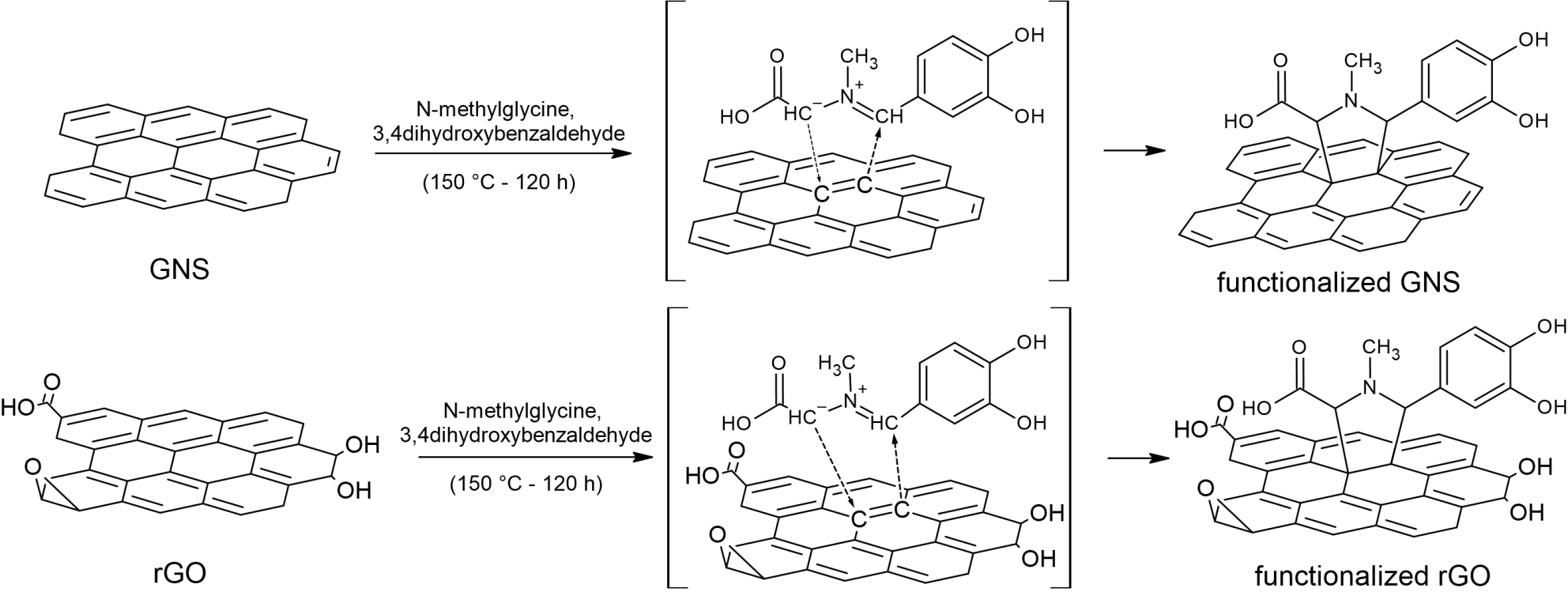}
		\caption{Schematic representation of the 1,3-DC of azomethine ylide on GNS and rGO.}
		\label{fig:DCscheme}
	\end{figure}
	
	\subsection{EDX/EELS analysis}
	
	\addcontentsline{toc}{subsection}{EDX/EELS analysis} 
	
	After the functionalization procedure, the dispersant solvent was removed by several washings with ethanol, and graphene was deposited on carbon film-supported copper TEM grids. Several scanning transmission electron microscope (STEM) images were acquired in order to identify the GNS (a representative STEM image is shown in Figure \ref{fig:OpticalRaman_EDX}(a)), while the successful functionalization of graphene was confirmed with energy-dispersive X-ray spectroscopy (EDX) and electron energy loss spectroscopy (EELS), which are very useful techniques for the elemental analysis or the chemical characterization of a sample. The elemental signature of nitrogen was detected mapping the functionalized GNS, and is shown as superimposed map (green signal) on the STEM image of the graphene flake in Figure \ref{fig:OpticalRaman_EDX}(b). It is interesting to notice that the molecules bond with graphene not only along the edges of the GNS, but also at the central C=C, as shown by the uniform distribution of the N signal in the EDX map (EDX and EELS spectra of functionalized GNS and rGO are available in the Supporting Information, Figures S2-S3).
	
	\begin{figure}[h]
		\centering
		\includegraphics[width=\linewidth]{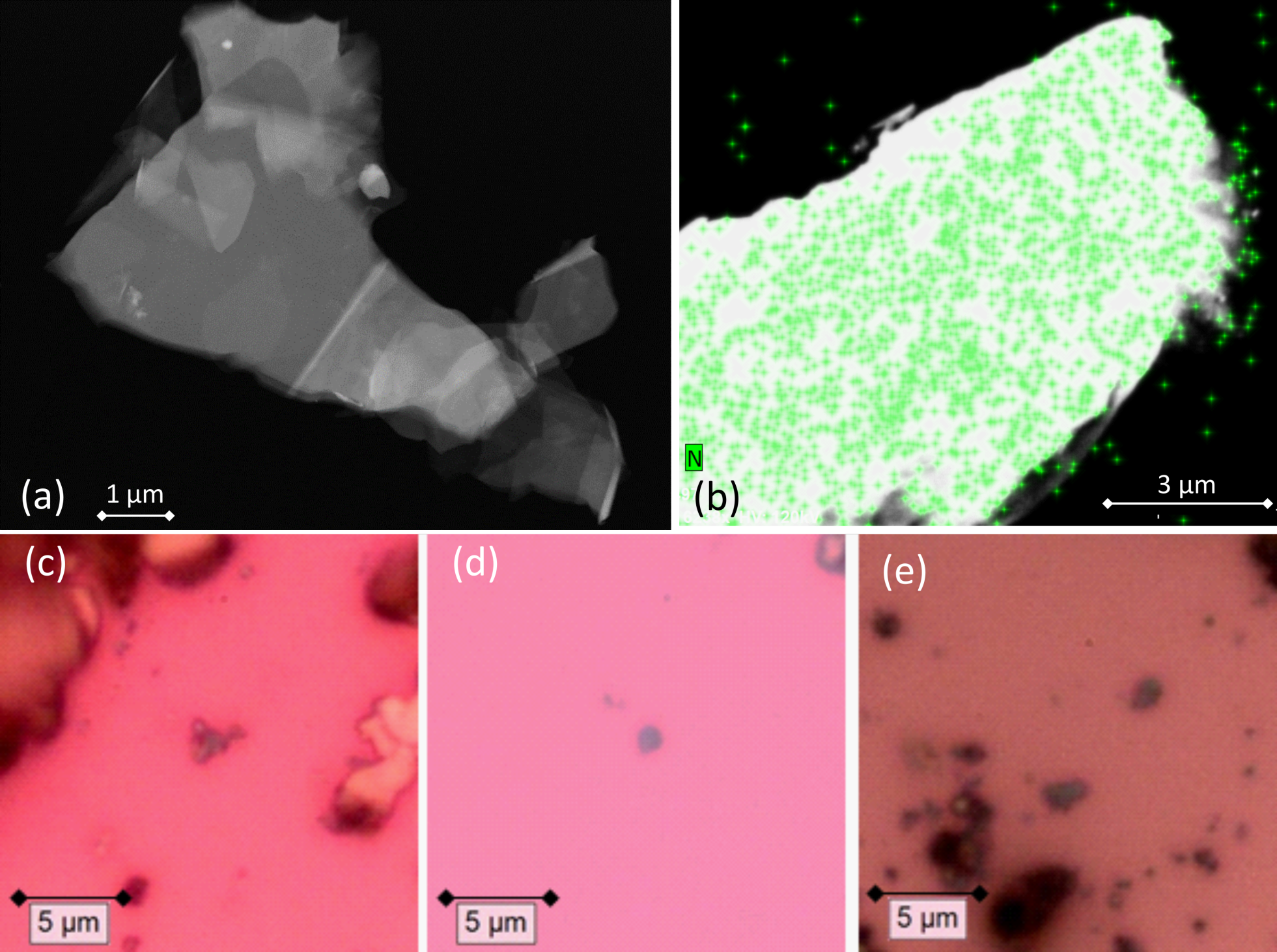}
		\caption{(a) STEM image of a functionalized GNS. (b) EDX map of a functionalized GNS, showing the uniform distribution of the N atoms (green pixels). (c-e) Optical microscopy images of functionalized GNS in (c) NMP and (d) in DMF, and (e) functionalized rGO in DMF dropcasted onto silica substrates for Raman measurements.}
		\label{fig:OpticalRaman_EDX}
	\end{figure}
	
	
	\subsection{Raman spectra of pristine and functionalized GNS}
	
	\addcontentsline{toc}{subsection}{Raman spectra of pristine and functionalized GNS}
	
	\begin{figure*}[t]
		\centering
		\includegraphics[width=0.9\linewidth]{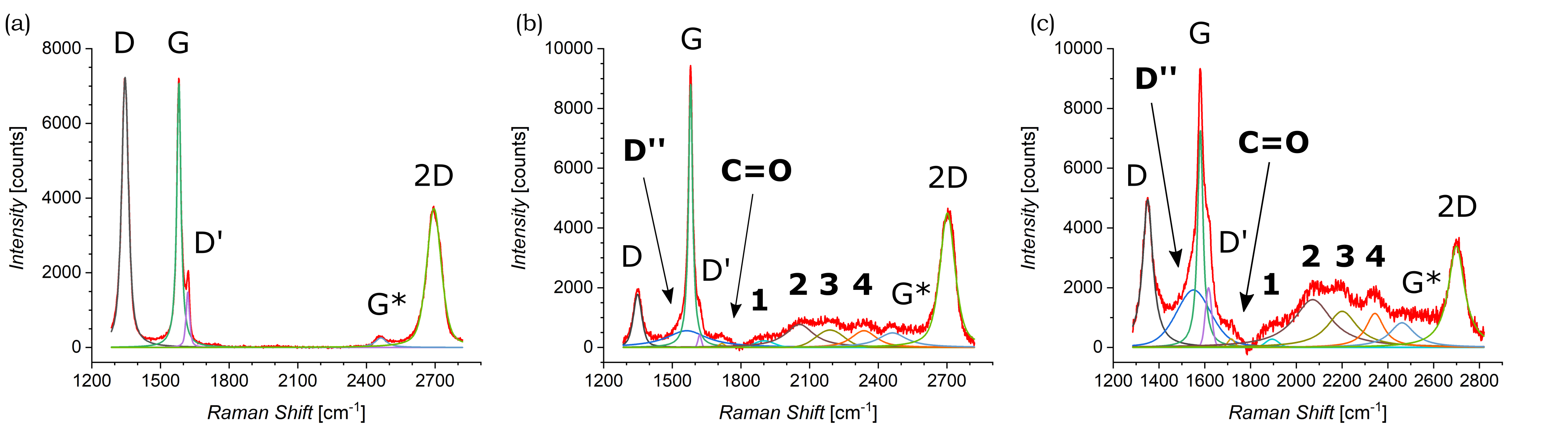}
		\caption{Raman spectra of (a) pristine GNS in NMP and (b) functionalized GNS in NMP and (c) in DMF. The fit of each spectrum is shown and all peaks are labeled (the peaks that appear only after the functionalization of GNS by 1,3-DC of azomethine ylide are in bold).}
		\label{fig:GNS_prepost}
	\end{figure*}
	
	Both pristine and functionalized GNS were dropcasted onto clean silica substrates, and the solvent was removed under vacuum (see Figure \ref{fig:OpticalRaman_EDX}(c,d)), in order to investigate them with Raman spectroscopy. Since a Raman spectrum of graphene was first recorded in 2006,\cite{Ferrari06} this technique has become one of the most suitable methods for the characterization of carbon-based nanomaterials, allowing both structural and electronic information, and being fast and non-destructive.\cite{Ferrari13} Figure \ref{fig:GNS_prepost}(a) shows the Raman spectrum of pristine GNS dispersed in NMP, along with the fit and the assignment of the 5 characteristic peaks of pristine graphene (the Raman spectrum of pristine GNS in DMF, very similar, is shown in the Supporting Information, Figure S5). The most recognizable features in the Raman spectrum of graphene are the G band, here centered at 1581 cm\textsuperscript{-1}, and the 2D band, here centered at 2702 cm\textsuperscript{-1}. The G band comes from the in-plane bond-stretching motion of pairs of C atoms of the graphene ring, while the 2D peak is the second order of the D peak and originates from a double-phonon process where momentum conservation is satisfied. Because no defects are required for their activation, both G and 2D peaks are always present. It is well-established that the width (FWHM) of the 2D band is related to the number of layers of graphene sheets.\cite{Ferrari06} In particular, a single curve with a FWHM of about 24~cm\textsuperscript{-1}\cite{Malard09} can be used for the fitting of the spectrum of monolayer graphene, while several curves (or a broader single one) are needed for the fitting of the spectrum of multilayer graphene. Here we can fit the 2D band with a single symmetrical Gaussian-Lorentzian curve, with a FWHM of about 70 cm\textsuperscript{-1}, which allows us to identify the GNS as few layer graphene (less than five layers thick, following the criterion described in Ref. \cite{Castillo18}). In the presence of small size graphene and/or defects, which we can refer to as a breaking of the symmetry of the infinite carbon honeycomb lattice,\cite{Eckmann12} additional bands appear. The D peak, here centered at 1347~cm\textsuperscript{-1}, and the D' peak, here centered at 1616~cm\textsuperscript{-1}, involve respectively intervalley and intravalley double resonance processes, and for low defect concentrations their intensities are proportional to the amount of defects.\cite{Wu18} The intensity ratio of the D and D' peaks has been shown to indicate the nature of the defects in the graphene lattice. In particular \textit{I}(D)/\textit{I}(D') reaches a maximum value of $\sim$ 13 for sp$^3$ defects, decreases to $\sim$ 7 for vacancy-like defects, and has a minimum of $\sim$ 3.5 for boundary-like defects.\cite{Eckmann12} Here, \textit{I}(D)/\textit{I}(D') is 4.89 in case of pristine GNS in NMP, and 4.02 in case of pristine GNS in DMF, confirming the boundary-like nature of the defects in pristine GNS. Finally, the weak G* band is visible, here centered at 2463 cm\textsuperscript{-1}, which arises from an intervalley process involving an in-plane transverse optical phonon and one longitudinal acoustic phonon (also called G + A\textsubscript{2U}).\cite{Krauss09}
	
	\begin{table*}[h]
		\small
		\caption{Presence (\textbullet) and position of the peaks from the fitting of the Raman spectra of pristine and functionalized GNS in NMP and in DMF.}
		\label{tab:GNS_peaks}
		\begin{tabular*}{0.98\textwidth}{@{\extracolsep{\fill}}cccccccccccc}
			\toprule
			GNS in NMP & D & D" & G & D' & C=O & 1 & 2 & 3 & 4 & G* & 2D \\
			\midrule
			pristine & \textbullet &  & \textbullet & \textbullet &  &  &  &  &  & \textbullet & \textbullet \\
			position [cm\textsuperscript{-1}] & 1346 &  & 1581 & 1620 &  &  &  &  &  & 2464 & 2695 \\
			error [cm\textsuperscript{-1}] & $\pm$ 1 &  & $\pm$ 1 & $\pm$ 1 &  &  &  &  &  & $\pm$ 2 & $\pm$ 1 \\   
			\midrule
			functionalized & \textbullet & \textbullet & \textbullet & \textbullet & \textbullet & \textbullet & \textbullet & \textbullet & \textbullet & \textbullet & \textbullet \\
			position [cm\textsuperscript{-1}] & 1349 & 1563 & 1580 & 1620 & 1717 & 1902 & 2057 & 2190 & 2338 & 2463 & 2702 \\
			error (cm\textsuperscript{-1}) & $\pm$ 1 & $\pm$ 3 & $\pm$ 1 & $\pm$ 2 & $\pm$ 2 & $\pm$ 6 & $\pm$ 10 & $\pm$ 10 & $\pm$ 5 & $\pm$ 3 & $\pm$ 1 \\
			\midrule
			GNS in DMF & D & D" & G & D' & C=O & 1 & 2 & 3 & 4 & G* & 2D \\
			\midrule
			pristine & \textbullet &  & \textbullet & \textbullet &  &  &  &  &  & \textbullet & \textbullet \\
			position [cm\textsuperscript{-1}] & 1346 &  & 1581 & 1620 &  &  &  &  &  & 2464 & 2695 \\
			error [cm\textsuperscript{-1}] & $\pm$ 1 &  & $\pm$ 1 & $\pm$ 1 &  &  &  &  &  & $\pm$ 2 & $\pm$ 1 \\
			\midrule
			functionalized & \textbullet & \textbullet & \textbullet & \textbullet & \textbullet & \textbullet & \textbullet & \textbullet & \textbullet & \textbullet & \textbullet \\
			position [cm\textsuperscript{-1}] & 1351 & 1552 & 1581 & 1617 & 1716 & 1895 & 2072 & 2201 & 2344 & 2463 & 2700 \\
			error [cm\textsuperscript{-1}] & $\pm$ 1 & $\pm$ 3 & $\pm$ 1 & $\pm$ 1 & $\pm$ 3 & $\pm$ 5 & $\pm$ 6 & $\pm$ 3 & $\pm$ 3 & $\pm$ 3 & $\pm$ 1 \\
			\bottomrule
		\end{tabular*}
	\end{table*}
	
	The Raman spectra of functionalized GNS exhibit new features and modifications, as shown in Figure~\ref{fig:GNS_prepost}(b,c). The intensity of the D peak substantially decreases. The ratio between D and G intensities is a benchmark for the grain size and the defect concentration.\cite{Ferrari07,Childres13} The ratio \textit{I}(D)/\textit{I}(G) passes from an initial value of 1.02 for pristine GNS in NMP to a final value of 0.20 for functionalized GNS in NMP. Similarly \textit{I}(D)/\textit{I}(G) decreases from 1.01 to 0.68 for GNS in DMF. The decrease of \textit{I}(D)/\textit{I}(G) can be explained considering that the azomethine ylides are grafting onto graphene's most favorable bonding sites, which are in the defected areas of the lattice, possibly leading to a local structural relaxation and a decrease in the Raman intensity of the defects. Considering that the GNS lateral size is comparable with the Raman laser spot ($\sim$~1~$\mu$m), the initial high intensity of the D peak before the functionalization also derives from the high density of edge defects, which are passivated by the presence of the azomethine ylides after the functionalization, leading to a decrease of the intensity of the D peak. Observing the 2D peak, it is noticeable that its intensity remains almost constant, with \textit{I}(2D)/\textit{I}(G) slightly changing from 0.53 to 0.51 for GNS in NMP and from 0.49 to 0.47 for GNS in DMF. This behavior of the 2D peak confirms that the long-range order of the graphene lattice is maintained, and that we are not introducing some strong disorder towards amorphous graphene. A further sign of the successful functionalization of GNS is the rise of a new broad peak centered at 1548~cm\textsuperscript{-1} for GNS in NMP and at 1545~cm\textsuperscript{-1} for GNS in DMF. This band, called D'' and usually seen in the range 1500 - 1550~cm\textsuperscript{-1}, is thought to be related to either the amorphous phase (increasing with the decrease of crystallinity)\cite{Vollebregt12} or to interstitial defects associated with the functionalization with small molecules.\cite{Sadezky05,Goodman13,Claramunt15} As aforementioned, the constant intensity of the 2D peak suggests that we are not inducing an amorphous phase, while the functionalization with azomethine ylide would explain creation of interstitial defects. Moreover, the peak centered at 1717~cm\textsuperscript{-1} for GNS in NMP and at 1716~cm\textsuperscript{-1} for GNS in DMF is usually assigned to the presence of the C=O functional group,\cite{Meade10} and would reasonably arise from the carboxyl group of the azomethine ylide. Finally, a set of four new peaks appears, which are labeled here as 1, 2, 3, and 4, centered respectively at around 1902 cm\textsuperscript{-1}, 2057~cm\textsuperscript{-1}, 2190~cm\textsuperscript{-1}, and 2338~cm\textsuperscript{-1} for GNS in NMP and at 1895~cm\textsuperscript{-1}, 2072~cm\textsuperscript{-1}, 2201~cm\textsuperscript{-1}, and 2343~cm\textsuperscript{-1} for GNS in DMF. These peaks are positioned in a region which is usually silent in Raman spectroscopy, and remarkably their positions do not significantly change with the change of the solvent (Figure S4 in the Supporting Information additionally shows that these peaks do not originate from the solvents). Hence, we tentatively attribute these peaks to Raman-active features of the azomethine ylide. This will be discussed in more detail in Section 1.6. Finally, the G* peak remains visible also in the spectra of functionalized GNS, centered at 2463 cm\textsuperscript{-1} for GNS in NMP and at 2459 cm\textsuperscript{-1} for GNS in DMF. Table~\ref{tab:GNS_peaks} summarizes the position and the appearance of all peaks, both for pristine and functionalized GNS in NMP and in DMF.
	
	\subsection{Raman spectra of pristine and functionalized rGO}
	
	\addcontentsline{toc}{subsection}{Raman spectra of pristine and functionalized rGO}
	
	Raman investigation was also performed on pristine and functionalized rGO, after dropcasting onto a silica substrate (see Figure \ref{fig:OpticalRaman_EDX}(e)) and drying under vacuum. Figure \ref{fig:rGO_prepost} shows the corresponding Raman spectra, which were collected in three overlapping regions (lower, intermediate, and higher) in order to allow a more detailed comparison. Because of the initial presence of oxygen functional groups in graphene oxide (GO) and the subsequent chemical reduction process, rGO presents several defects in its structure.\cite{Stankovich07,Bagri10} This leads to the broadening of the characteristic D and G bands of graphene, here centered at 1345~cm\textsuperscript{-1} and 1595~cm\textsuperscript{-1}, as shown in Figure \ref{fig:rGO_prepost}(a). Of fundamental interest is the broad valley between the two peaks, which has been reported for carbon-based materials and rGO.\cite{Sadezky05,Claramunt15,Cuesta94} By deconvolution we can identify three additional peaks, centered at 1240~cm\textsuperscript{-1}, 1515~cm\textsuperscript{-1}, and 1621~cm\textsuperscript{-1}. The first peak, usually called D*, can be related to disordered graphitic lattices provided by sp\textsuperscript{2}-sp\textsuperscript{3} bonds at the edges of networks,\cite{Sadezky05} whereas
	the second peak can be identified as the D'' band and the last one as the D' band. While in pristine GNS the D'' peak is not present, here its appearance can be explained by the presence of a number of C sp\textsuperscript{3} bonds from residual functional groups already present in the rGO structure before the functionalization. Correspondingly, the band centered at 1734~cm\textsuperscript{-1} can be attributed to the presence of C=O functional groups. Moreover, a weak band centered at 1108~cm\textsuperscript{-1} is visible. A similar peak around 1150~cm\textsuperscript{-1} has been reported in nanocrystalline diamond to be due to the sum and difference modes of sp\textsuperscript{2} C=C and C-H vibrations of trans-polyacetylene-type segments occurring at grain boundaries.\cite{Ferrari01} Likewise, a peak around 1130~cm\textsuperscript{-1} has been observed arising from the edges and holes of GO flakes which present similar H-ending C=C chains.\cite{Diez13} We use here the same nomenclature, labeling it as sp\textsuperscript{2}-bound. As expected, in higher disorder graphene\cite{Ferreira10} the intensity of the 2D band decreases and broadens, as observed in Figure \ref{fig:rGO_prepost}(b). At higher Raman shift values (see Figure \ref{fig:rGO_prepost}(c)), besides the 2D band centered at 2681~cm\textsuperscript{-1}, the D+D' combination band is visible, centered at 2933~cm\textsuperscript{-1}. Finally, we can assign the peak centered at 3179 cm\textsuperscript{-1}, to the C-H stretching mode.\cite{Diez13}

	\begin{figure*}[h]
		\centering
		\includegraphics[width=0.9\linewidth]{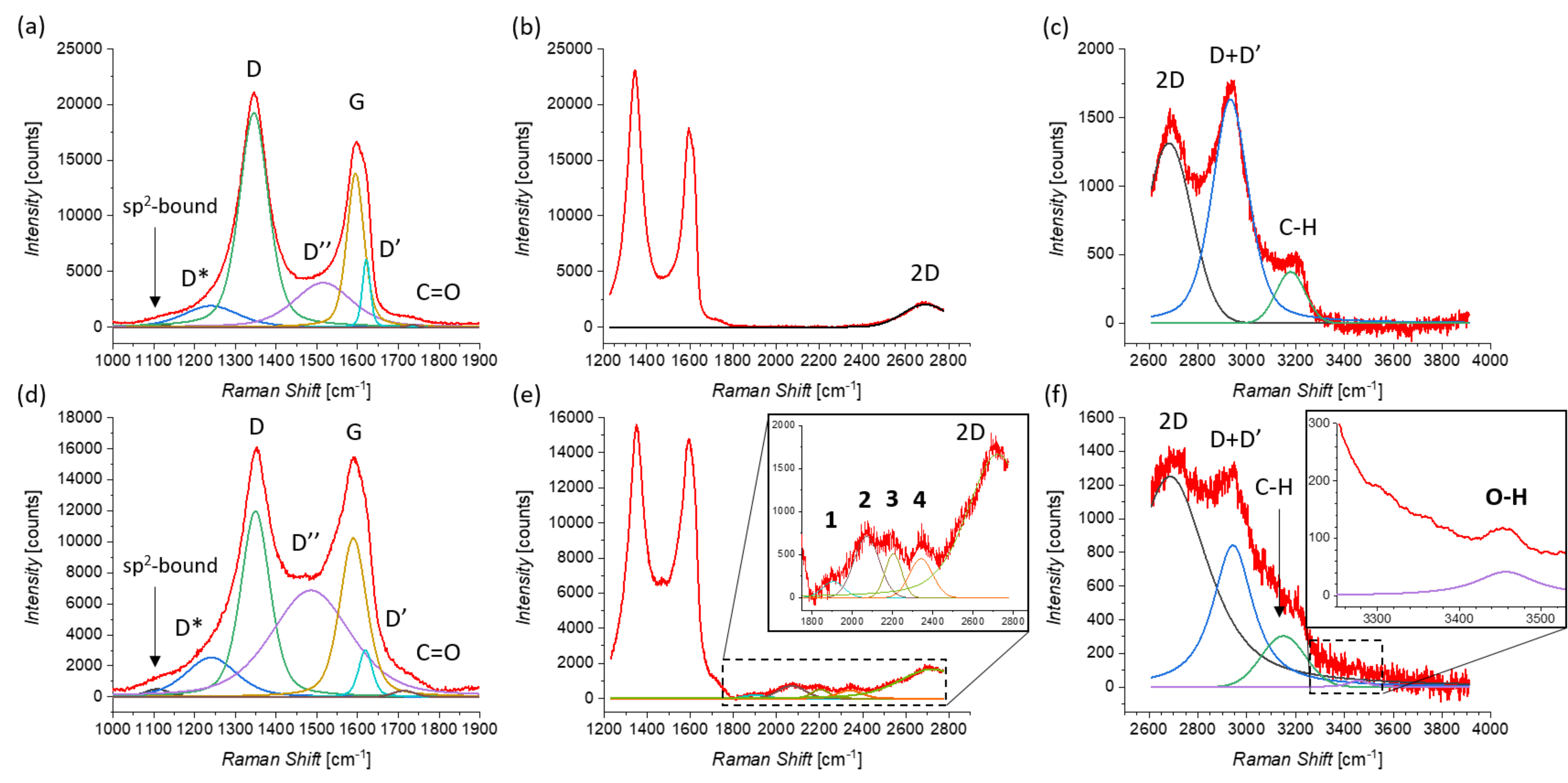}
		\caption{Raman spectra of pristine rGO taken in the (a) lower, (b) intermediate, and (c) higher region of Raman shifts. Raman spectra of functionalized rGO taken in the (d) lower, (e) intermediate (zoom is shown in the inset), and (f) higher region (zoom, where a smoothing of the signal was performed, is shown in the inset) of Raman shifts. The fit of each spectrum is shown and all peaks are labeled (the peaks that appear only after the functionalization of rGO by 1,3-DC of azomethine ylide are in bold).}
		\label{fig:rGO_prepost}
	\end{figure*}
	
	The Raman spectrum of functionalized rGO shows new and interesting features. As already seen for functionalized GNS, the intensity of the D peak decreases (as shown in Figure \ref{fig:rGO_prepost}(d)) and \textit{I}(D)/\textit{I}(G) passes from 1.39 to 1.17 after the 1,3-DC of azomethine ylide. Likewise, \textit{I}(D')/\textit{I}(G) decreases from 0.44 to 0.29. Furthermore, the intensity of the D'' peak increases, with \textit{I}(D'')/\textit{I}(G) moving from 0.29 before to 0.67 after the 1,3-DC, indicating the functionalization due to the presence of azomethine ylides. It is interesting to notice that also the intensity of the Raman signal from the C=O group increases. As shown in Figure \ref{fig:rGO_prepost}(e), the same set of four peaks (1 - 4) arises in the Raman spectrum after the functionalization. Similarly to what we see in functionalized GNS, peak 1 is centered at 1904 cm\textsuperscript{-1}, peak 2 at 2071~cm\textsuperscript{-1}, peak 3 at 2204 cm\textsuperscript{-1}, and peak 4 at 2341 cm\textsuperscript{-1}. In the region of higher Raman shift, shown in Figure \ref{fig:rGO_prepost}(f), the decrease of the intensity of the D+D' peak is clearly visible, explicable with the corresponding decrease of the D and D' intensities. Finally, a new weak band centered at 3457 cm\textsuperscript{-1} appears. This peak can be attributed to the O-H stretching\cite{Kawamoto04} of the carboxyl or the catechol groups in the azomethine ylide, and to our knowledge this is the first time it is seen in functionalized graphene. Table \ref{tab:rGO_peaks} summarizes the position and the appearance of all peaks, for both pristine and functionalized rGO.
	
	
	Significantly, the appearance of the peaks 1 - 4 is in complete agreement between functionalized GNS in NMP and in DMF and functionalized rGO in DMF. Moreover, the exact correspondence of the positions of the peaks 1 - 4 within error intervals is noteworthy. This result corroborates the conclusion that these peaks do not originate from the solvents (same position in NMP and in DMF) or the substrate (same position for GNS and rGO), but arise from the functionalization with the azomethine ylide.

	\begin{table*}[h]
		\footnotesize
		\caption{Presence (\textbullet) and position of the peaks from the fitting of the Raman spectra of pristine and functionalized rGO in DMF.}
		\label{tab:rGO_peaks}
		\begin{tabular*}{\textwidth}{@{\extracolsep{\fill}}cccccccccccccccc}
			\toprule
			rGO in DMF & sp\textsuperscript{2}-bound & D* & D & D" & G & D' & C=O & 1 & 2 & 3 & 4 & 2D & D+D' & C-H & O-H \\
			\midrule
			pristine & \textbullet & \textbullet & \textbullet & \textbullet & \textbullet & \textbullet & \textbullet &  &  &  &  & \textbullet & \textbullet & \textbullet & \\
			position [cm\textsuperscript{-1}] & 1108 & 1240 & 1345 & 1515 & 1595 & 1621 & 1722 &  &  &  &  & 2681 & 2933 & 3179 &  \\
			error [cm\textsuperscript{-1}] & $\pm$ 10 & $\pm$ 2 & $\pm$ 1 & $\pm$ 3 & $\pm$ 1 & $\pm$ 2 & $\pm$ 5 &  &  &  &  & $\pm$ 2 & $\pm$ 1 & $\pm$ 2 &  \\
			\midrule
			functionalized & \textbullet & \textbullet & \textbullet & \textbullet & \textbullet & \textbullet & \textbullet & \textbullet & \textbullet & \textbullet & \textbullet & \textbullet & \textbullet & \textbullet & \textbullet \\
			position [cm\textsuperscript{-1}] & 1108 & 1241 & 1349 & 1486 & 1589 & 1619 & 1714 & 1904 & 2071 & 2204 & 2341 & 2686 & 2943 & 3149 & 3457 \\
			error [cm\textsuperscript{-1}] & $\pm$ 4 & $\pm$ 2 & $\pm$ 1 & $\pm$ 2 & $\pm$ 1 & $\pm$ 2 & $\pm$ 4 & $\pm$ 5 & $\pm$ 6 & $\pm$ 4 & $\pm$ 3 & $\pm$ 1 & $\pm$ 1 & $\pm$ 3 & $\pm$ 5 \\
			\bottomrule
		\end{tabular*}
	\end{table*}
	
	\subsection{Computational simulations}
	
	\addcontentsline{toc}{subsection}{Computational simulations}
	
	In order to deepen our understanding of the functionalized graphene after 1,3-DC of azomethine ylide, models for pristine and functionalized rGO were built. Pristine rGO was modeled by positioning 10 oxygen atoms on top of randomly chosen carbon atoms, followed by structural minimization (as described in detail in Section 3). The total amount of oxygen was 8\% in weight, in agreement with similar rGO models in literature.\cite{Bagri10} Functionalized rGO was modeled by adding the azomethine ylide in the central area of the rGO structure, followed by structural minimization. Functionalized rGO appears mildly corrugate, with a final morphology depending on the position of the epoxy groups and the azomethine ylide.
	
	Both models were characterized by evaluating the restrained electrostatic potential (RESP) derived partial atomic charges.\cite{Bayly93} The RESP-derived partial charges, mapping the electrostatic features of the system, highlight the localization of the charges in the pristine rGO induced by the presence of the epoxy groups. The highest positive/negative charge values are, in fact, localized in the sp$^3$ carbon atoms linked to the epoxy groups (blue/red in Figure \ref{fig:SimMap}(a)). The partial charge in the sp$^3$ atoms affects theirs neighbors, an effect that gradually decreases at longer distances. Remarkably, the presence of the azomethine ylide amplifies the localization ot the charges (see Figure \ref{fig:SimMap}(b)), demonstrating that the functionalization can modulate the distribution of charges on atoms, eventually favoring the further binding of additional molecules.
	
	\begin{figure}[h]
		\centering
		\includegraphics[width=\linewidth]{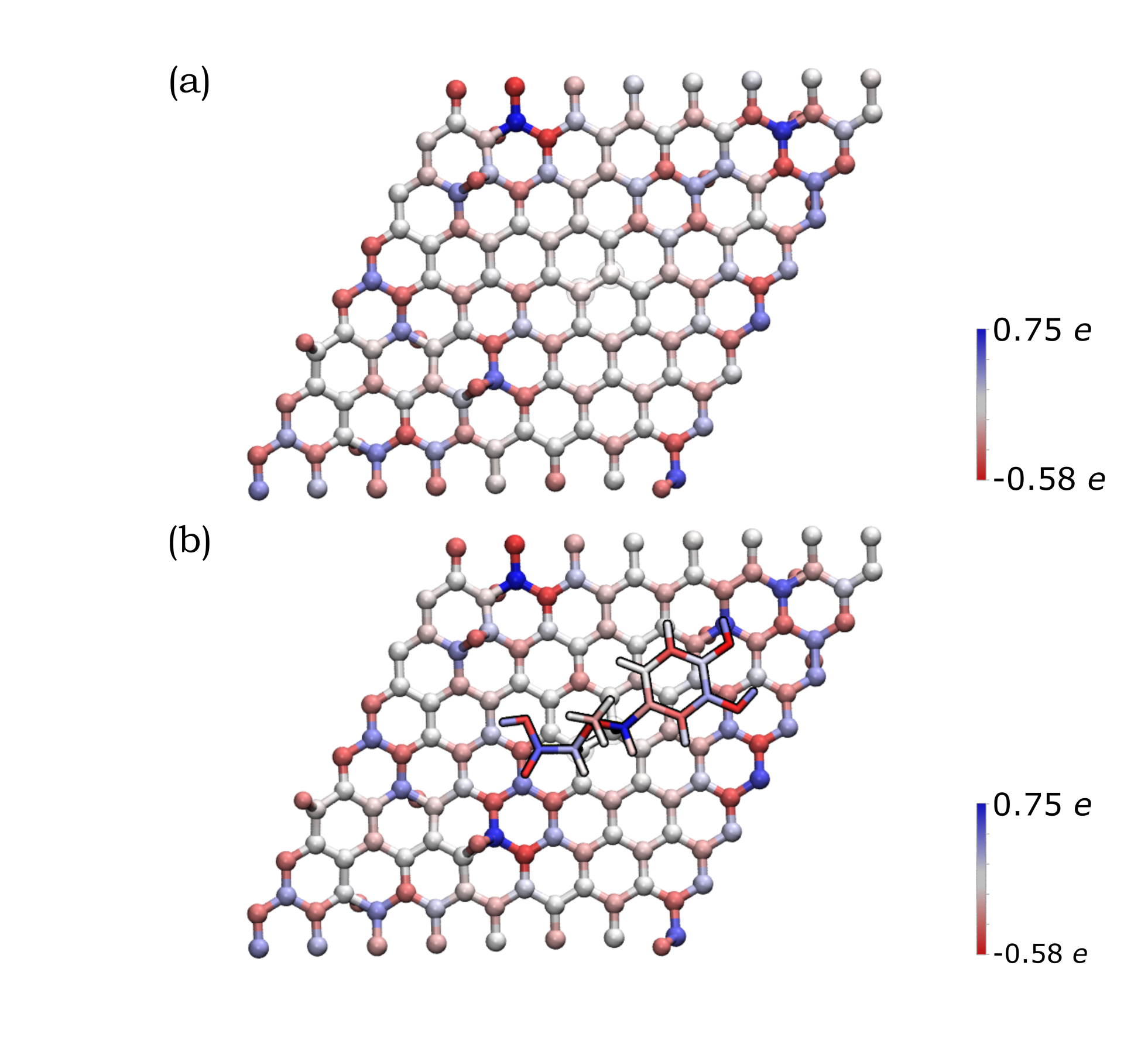}
		\caption{Distribution of the RESP-derived partial atomic charges reported as a color gradient for the atoms in (a) pristine rGO and (b) functionalized rGO (blue = positive, red = negative). The highest positive/negative charge values are close to the epoxy groups (that present O atoms outside the graphene plane). The amplification of the charge gradient due to the presence of the azomethine ylide is visible.}
		\label{fig:SimMap}
	\end{figure}
	
	The functionalized system was then subjected to 12 ps of molecular dynamics (MD) simulation at 300~K. During the MD simulation, the lattice did not undergo a large conformational rearrangement. The vibrational density of states has been obtained as the Fourier transform of the velocity autocorrelation function. The power spectrum was then decomposed and analyzed by computing the power spectra of the autocorrelation function of appropriate groups of atomic coordinates (highlighted in different colors in Figure \ref{fig:SimSpectrum}(a)), and is shown in Figure \ref{fig:SimSpectrum}(b). In the intermediate region, the characteristic vibrational peak from the C=O in the carboxyl group of the azomethine ylide is visible, centered at 1730~cm\textsuperscript{-1}. Moreover, there is a set of four vibrational regions (labeled as 1', 2', 3', and 4' in the inset of Figure \ref{fig:SimSpectrum}(b)) corresponding to the C-N stretching from the nitrogen of the azomethine ylide, centered around 920~cm\textsuperscript{-1}, 980~cm\textsuperscript{-1}, 1050~cm\textsuperscript{-1}, and 1200~cm\textsuperscript{-1} respectively. If we consider the second order of these bands, we obtain four vibrational bands around (1900 $\pm$ 100)~cm\textsuperscript{-1}, (2000 $\pm$ 100)~cm\textsuperscript{-1}, (2100 $\pm$ 100)~cm\textsuperscript{-1}, and (2400 $\pm$ 100)~cm\textsuperscript{-1} (the errors are related both to the simulation and to the extension of the regions 1' - 4'). These values are consistent with the Raman features which appear in the intermediate region of both functionalized GNS and rGO. The corresponding first-order Raman bands are not visible in the experimental Raman spectra, being covered by the broad shoulder from the silicon substrate around 920 - 1050~cm\textsuperscript{-1} (shown in Figures S6-S7 in the Supporting Information). The simulated power spectrum also exhibits a vibrational band from the C-H groups of the aromatic catechol of the azomethine ylide around 2980 - 3200~cm\textsuperscript{-1}, in agreement with the experimental peak seen in functionalized rGO. Finally, the O-H groups of the ylide possess stretching bands around 3460~cm\textsuperscript{-1} and over 3650 cm\textsuperscript{-1}, in agreement with the result from the experimental Raman spectrum of functionalized rGO.
	
	\begin{figure}[h]
		\centering
		\includegraphics[width=\linewidth]{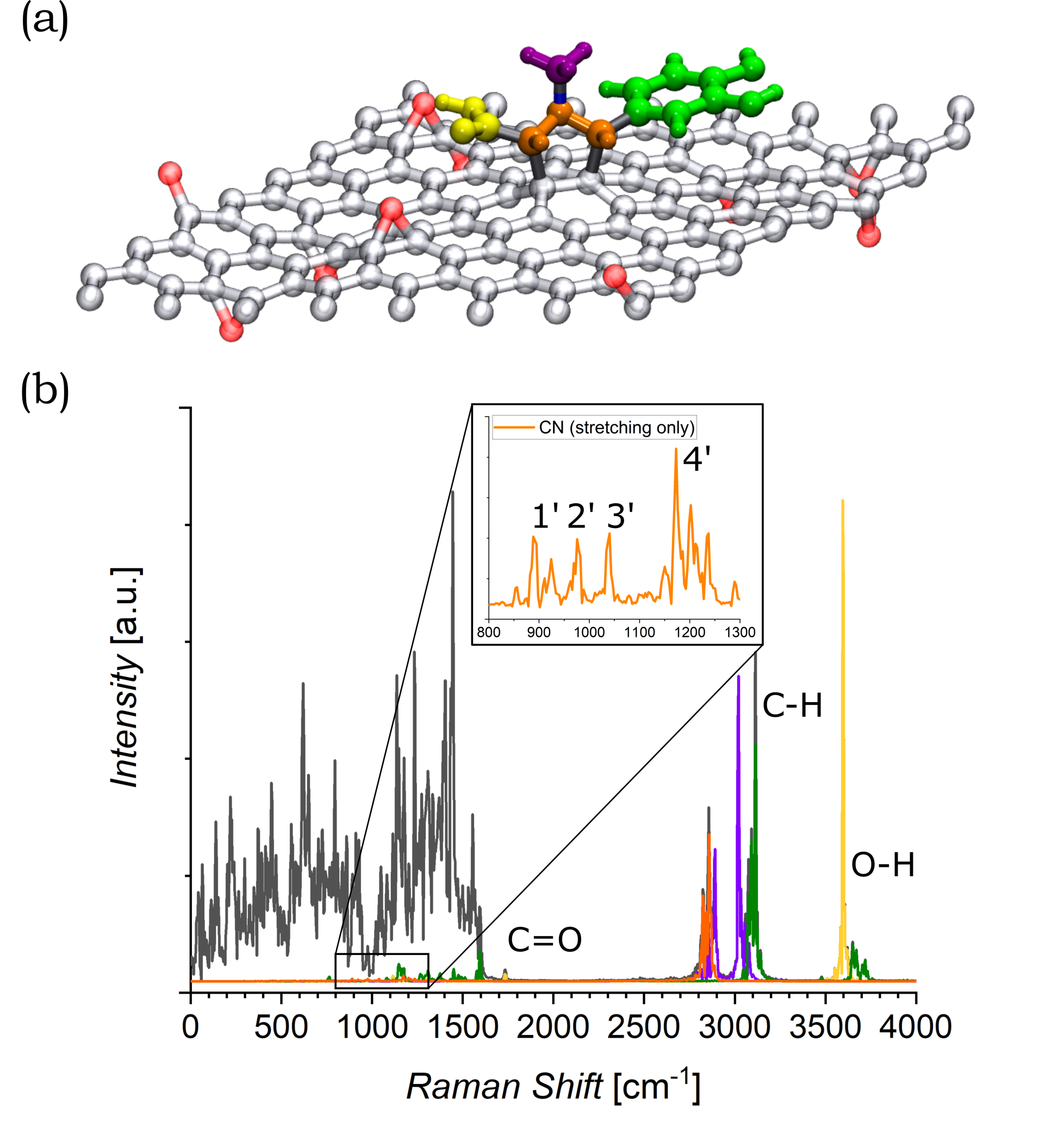}
		\caption{(a) Model of the azomethine ylide attached to rGO, where the red atoms are the oxygens of the epoxy groups. The functional groups of interest are highlighted: carboxyl group = yellow, nitrogen and the connected atoms = orange, methyl group = violet, cathecol group = green. (b) Power spectrum (PS) of the velocity autocorrelation function (black). Projections on the PS for the functional groups of interest are highlighted in the same colors of panel (a). The regions of interest are labeled. The inset shows the zoom of the region where the stretching of the CN bonds are visible. It is worth to recall that the intensities of the peaks in the simulated PS do not directly correspond to the ones from the experimental Raman spectra, where additional selection rules are involved.}
		\label{fig:SimSpectrum}
	\end{figure}

	\subsection{XPS analysis}
	
	\addcontentsline{toc}{subsection}{XPS analysis}
	
	Pristine and functionalized GNS and rGO were dropcasted onto silica substrates, and the solvent was removed under vacuum. This procedure was repeated several times in order to achieve a homogeneous coverage of the entire surface of the substrate, allowing measurements without any signal arising from the substrate. X-ray photoelectron spectroscopy provides a powerful and direct method for the determination of the surface elemental composition of a material. This information is crucial in order to quantify the efficiency of the functionalization, as well as the atomic abundance. In this study, the investigation was focused on the C 1s, N 1s, and O 1s core levels, which are related to the graphene substrate and the functionalizing material. Performing the deconvolution of the C 1s core level photoemission spectra of functionalized graphene (both GNS and rGO), we can identify five different components Figure~\ref{fig:XPS_DMF_C1s} shows the spectrum of functionalized GNS in DMF, while the other spectra are available in the Supporting Information, Figures S8-S9). The main peak centered at a binding energy (B.E.) of 284.4 eV can be assigned to the sp\textsuperscript{2} C-C bonds of the graphene sheet, while a second weaker peak due to sp\textsuperscript{3} C-C bonds is distinguishable at 285.2 eV.\cite{Konicki17} The additional peak arising from carbon atoms bound to oxygen (C-O) and to nitrogen (C-N) are found at 286.1 eV.\cite{Benne06} Finally, the contributions from C=O and O-C=O are seen at 287.8 eV and 290.8~eV, respectively.\cite{Xing16}
	
	\begin{figure}[h]
		\centering
		\includegraphics[width=\linewidth]{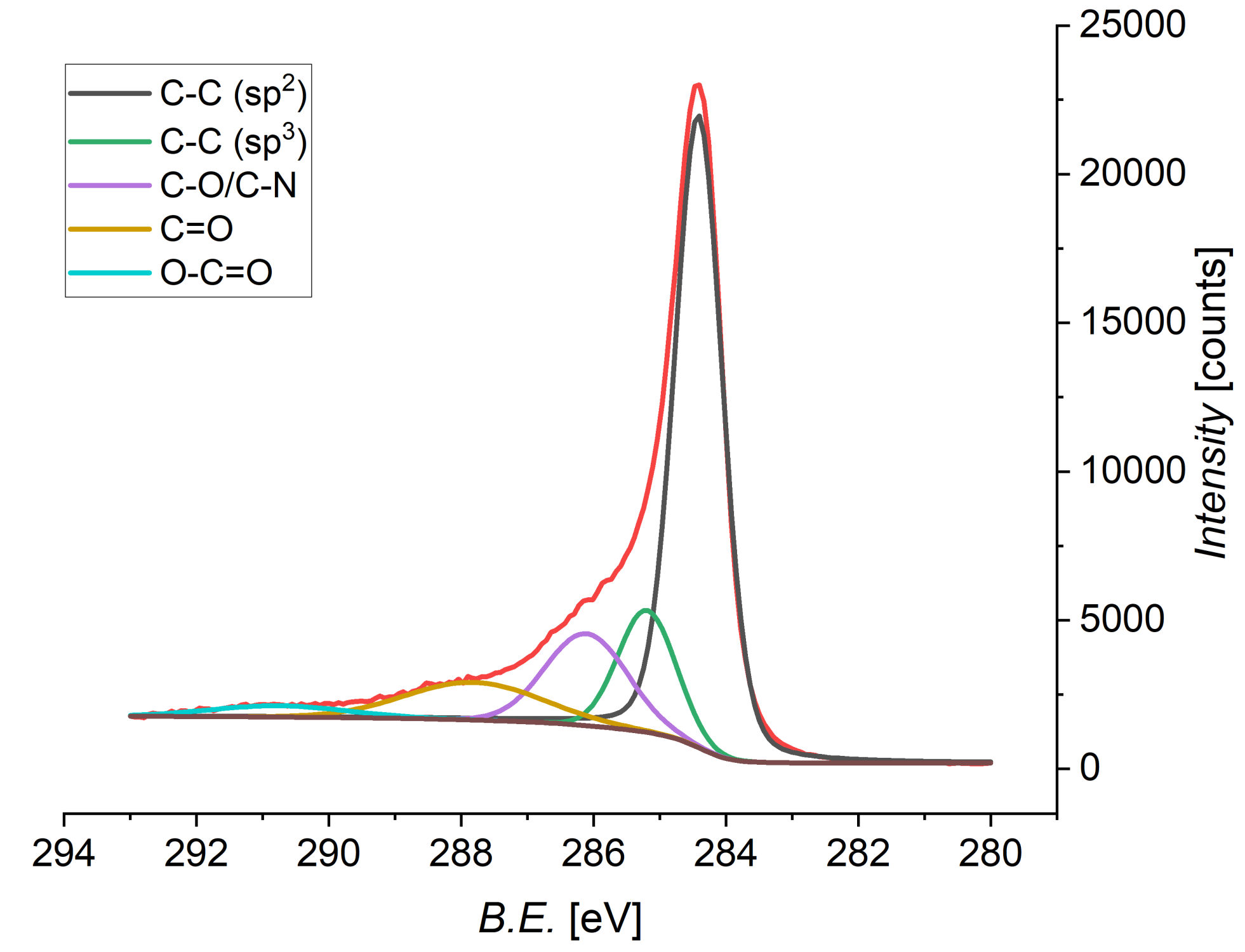}
		\caption{XPS spectrum of the C 1s core level of functionalized GNS in DMF, and the fit showing the individual components (Shirley-type background in brown).}
		\label{fig:XPS_DMF_C1s}
	\end{figure}
	
	Further information can be derived from the N 1s core level photoemission spectra of functionalized graphene Figure \ref{fig:XPS_N1s}). All spectra display a shoulder for lower values of the B.E. which can be assigned to the peak coming from the N atom of the azomethine ylide. This peak is always centered at around 398.8~eV. The principal peaks seen in the N 1s spectra are identified as coming from the N atoms of the solvents. As shown in Figure S11 in the Supporting Information, the spectrum of pristine GNS in NMP is peaked at 400.9 eV, while the spectra of pristine rGO in DMF is peaked at 400.0 eV. These components arise from the residual solvent that remains adsorbed or trapped after the dropcasting process and was not removed during the drying under vacuum. This issue is therefore related to the specific sample preparation procedure for XPS. Indeed, Raman analysis has never shown peaks related to the solvent. After annealing of functionalized GNS in DMF at 90 °C, 130 °C, and 180 °C the intensity of the peak deriving from the solvent decreases, indicating its desorption (as shown in Figure S12 in the Supporting Information). Remarkably, the intensity of the peak arising from the azomethine ylide remains almost constant, validating the stability of the functionalization. The isolated peak centered at 403.7~eV, visible only in functionalized GNS in NMP, derives from the NO\textsubscript{2} (from partial degradation of the NMP solvent) bonded to Na (added to lower the acidity of the environment, as explained in Section 3), and disappears after annealing at 180 °C.\cite{Datta84} From the intensities of the C 1s, N 1s, and O~1s spectra, we can calculate the elemental abundances after the 1,3-DC of azomethine ylide, which are presented in Table \ref{tab:atomic_percentage} (the O 1s spectra are shown in the Supporting Information, Figure S10). As expected, the oxygen abundance in functionalized rGO is higher than in functionalized GNS, deriving from the fact that pristine rGO already contains around 13\% of O (data from the supplier). The data in Table \ref{tab:atomic_percentage} also provide an estimation of the efficiency of the functionalization, resulting in the presence of 1 azomethine ylide every $\sim$ 225 carbons in case of GNS in NMP, 1 ylide every $\sim$ 170 carbons for GNS in DMF, and 1 ylide every $\sim$ 110 carbons for rGO in DMF. A higher degree of functionalization for GNS in DMF is consistent with the fact that DMF is a better organic solvent for the reaction process, having a superior capability to stabilize the reaction intermediate, with respect to NMP. The higher coverage of rGO is attributed to the higher presence of defects in its graphene-like surface, which increases the number of favorable binding positions for the azomethine ylide. As shown in the RESP-derived charge map, the epoxy groups in the pristine structure affect the electronic distribution, providing localized partial charges in the pristine rGO which favor the 1,3-DC.
	
	\begin{figure}[h]
		\centering
		\includegraphics[width=\linewidth]{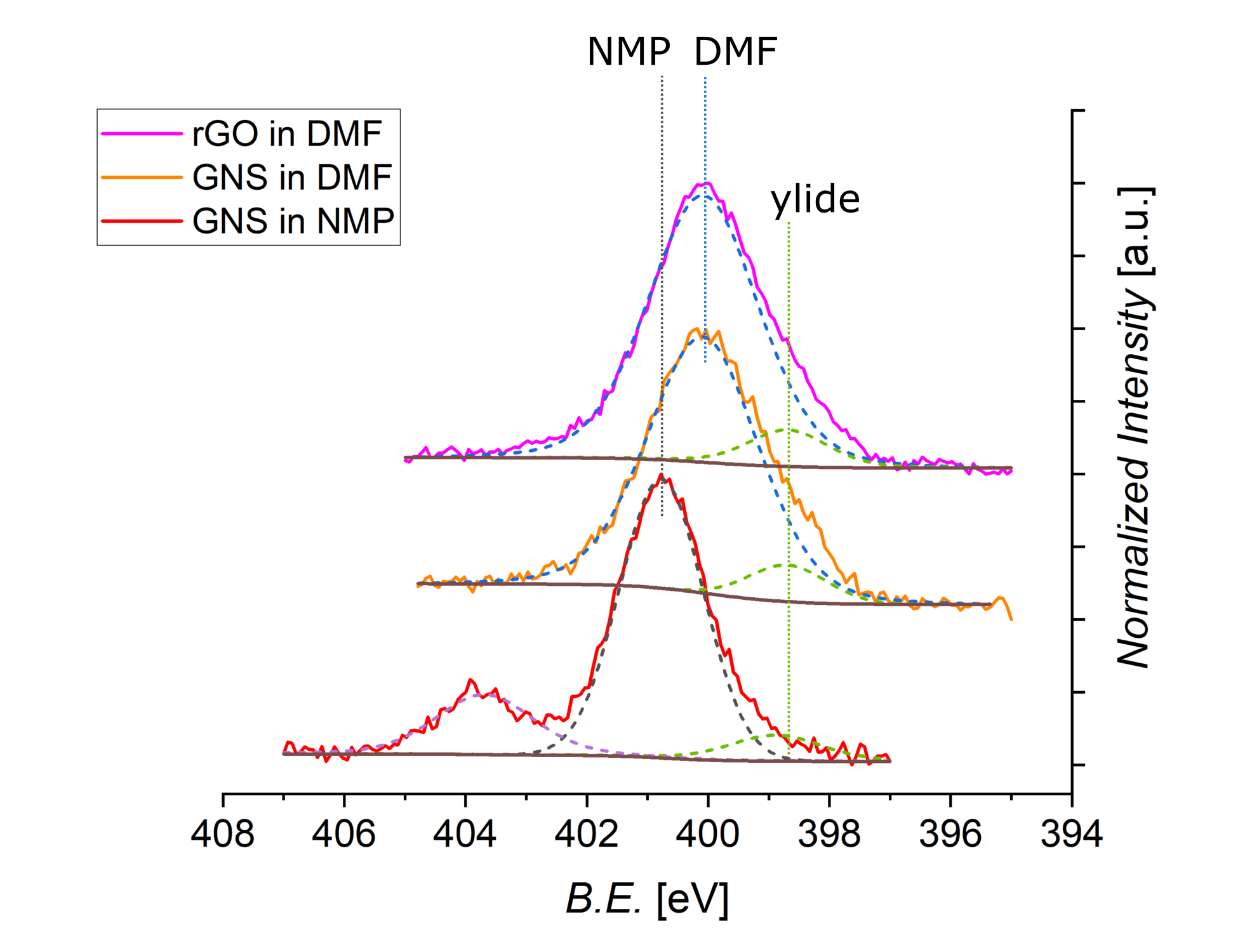}
		\caption{Normalized XPS spectra of N 1s core levels of functionalized GNS in NMP, GNS in DMF, and rGO in DMF (shifted in height). The different positions of the peak related to the solvent are labeled (dashed black line for NMP, dashed blue line for DMF), together with the constant position of the peak from the ylide (dashed green line). Each fit is shown by a dashed line, with a Shirley-type background in brown (the spectrum of functionalized GNS in NMP shows an additional signal arising from NaNO\textsubscript{2}, and its fit is shown as dashed violet line).}
		\label{fig:XPS_N1s}
	\end{figure}
	
	\begin{table}[h]
		\small
		\caption{Elemental composition of functionalized GNS and rGO, calculated from the XPS intensities.}
		\label{tab:atomic_percentage}
		\begin{tabular*}{0.48\textwidth}{@{\extracolsep{\fill}}ccccc}
			\hline
			& C [\%] & N(ylide) [\%] & N(solvent) [\%] & O [\%] \\
			\hline
			GNS in NMP  & 80.2  & 0.34 & 3.7 & 15.7 \\
			GNS in DMF  & 82.1 & 0.45 & 4.6 & 12.8 \\
			rGO in DMF  & 72.6 & 0.60 & 5.4 & 21.4 \\
			\hline
		\end{tabular*}
	\end{table}
	
	
	\section{Conclusions}
	
	\addcontentsline{toc}{section}{Conclusions} 
	
	In this work we have successfully functionalized graphene via 1,3-dipolar cycloaddition of azomethine ylide. For the first time, a comparison of the efficiency of this reaction in different dispersant solvents is provided. Even if the dispersion capability of NMP for graphene is slightly higher, the reaction proceeds more efficiently in DMF, allowing a greater organic functionalization of graphene in the liquid phase. To begin with, the functionalization was confirmed by detecting the presence of nitrogen in functionalized graphene with XPS, EDX, and EELS measurements. New Raman features arising from the functionalization with azomethine ylide were detected both in functionalized GNS and in functionalized rGO, and were assigned with the help of DFT simulations of the vibrational power spectrum of functionalized graphene. Thanks to the local inhomogeneity of the partial charges, due to the presence of oxygen functional groups in the pristine structure, a higher functionalization was achieved on rGO. This validates the interest in further exploring the possibility to control the position and the nature of defects, towards a precise building up of specific nanostructures, by exploring the design of the organic functionalization. Further chemical modifications of the functionalized graphene could be implemented by activating the carboxyl group of the azomethine ylide, with N-hydroxysuccinimmide for example, which would allow the binding of linker molecules and additional functional groups. Moreover, the efficiency of 1,3-DC of azomethine ylide suggests its use in future organic functionalization of higher quality graphene systems, like exfoliated flakes or epitaxial monolayer samples, where the controlled use of defect engineering would allow the realization of tailored devices starting from a performing substrate, in terms of transport properties. These results open the route for a wider range of applications, such as electrochemical devices for gas sensing or storage, enzymatic essays, catalysis, or protein interaction.


\section{Methods}

\addcontentsline{toc}{section}{Methods} 

\subsection{Chemicals}
Wet-jet milling exfoliated graphene powder was provided by BeDimensional Srl, Italy. Reduced graphene oxide (rGO) in flake form was purchased from Graphenea, Spain. 1-methyl-2-pyr\-rol\-id\-i\-none (ReagentPlus, 99\%), N,N-dimethylformamide (anhydrous, 99.8\%), ethanol (puriss, $\geq$96\%), N-methylglycine (98\%), and 3,4-dihydroxy\-ben\-zalde\-hyde (97\%) were purchased from Sigma-Aldrich/Merck, Germany, while sodium carbonate (for analysis, $\geq$99.5\%) was purchased from Carlo Erba, Italy.

\subsection{Dispersion of GNS and rGO}
Graphene powder (35~mg) was dispersed in NMP or DMF (140~mL). For the dynamic light scattering (DLS) measurements, homogenization (OV5 Homogenizer from VELP Scientifica) was performed in time steps of 10 minutes, up to 1 hour, at 30000~rpm. Sonication (SONOPLUS Ultrasonic HD 2070 from Bandelin) was performed in different time steps, from 15 minutes to 2 hours, at 60~W. For the functionalization procedure, homogenization was performed for 30 minutes at 30000~rpm. Reduced graphene oxide powder (10~mg) was dispersed in DMF (40~mL). Homogenization was then performed for 30 minutes at 30000~rpm. 

\subsection{1,3-DC of azomethine ylide on GNS or rGO}
To perform the organic functionalization, 3,4-\-di\-hy\-drox\-y\-ben\-zal\-de\-hyde (25~mg, 0.18~mmol) and an excess of N-methylglycine (25~mg, 0.28~mmol)  were added to the graphene dispersion (22~mL, 0.20~mg~mL\textsuperscript{-1}) in NMP or DMF, similarly for rGO dispersed in DMF (22~mL, 0.20~mg~mL\textsuperscript{-1}). Sodium carbonate (5~mg, 0.05~mmol) was added in order to lower the acidity of the environment. This both helps the formation of the 1,3-dipolar compound and prevents the loss of the carboxyl group of the azomethine ylide. The reaction mixture was heated at 150~°C for 120~h, under magnetic stirring, with successive additions of the reagents every 24~h. In order to limit secondary reactions from the oxidation of the solvent at high temperature, an inert atmosphere (N\textsubscript{2}) was kept during the functionalization reaction. The resulting mixture was then thoroughly washed with clean solvent (several passages of centrifugation and pipetting off the solvent).

\subsection{Characterization Techniques}
DLS measurements were performed with a Malvern Zetasizer Nano ZS (equipped with a 633 nm HeNe laser), using quartz cuvettes (1 mL). Samples were equilibrated to 25~°C for 30 seconds before measuring and the data were acquired with a scattering angle of 12.8°. Values for NMP viscosity and refractive index at 25~°C are 1.6660 cP and 1.468, while for DMF are 0.8020~cP and 1.428 (data from the supplier). Graphene refractive index and absorption at 25 °C are 2.704 and 0.023,\cite{Weber10} respectively. Raman spectroscopy was carried out with a Renishaw InVia system, equipped with a confocal microscope, a 532 nm excitation laser and a 1800 line/mm grating (spectral resolution 2 cm\textsuperscript{-1}). All spectra were measured with the following parameters: excitation laser power 500 $\mu$W, acquisition time for each spectrum 20~s, two acquisitions per spectrum, with a 100X objective (NA=0.85, spot size 1~$\mu$m). At least 3 different samples were measured, both for pristine and functionalized graphene, and they all showed the same features of the ones here reported as representative results. The samples were prepared by dropcasting (2~$\mu$L) onto clean silica substrate and dried under vacuum for several days. STEM–EDX/EELS measurements were carried out on a Zeiss Libra 120 microscope operating at an accelerating voltage of 120~kV, equipped with a thermionic LaB\textsubscript{6} source and an in-column Omega filter for energy-filtered imaging and EELS analysis. EELS spectra were collected by integrating the EELS signal on the TEM CCD (a TRS 2k x 2k CCD camera binned 2 by 2) and by selecting the measured area using the entrance slit aperture of the Omega filter. The samples were prepared by washing the dispersant solvent with ethanol several times and collected with carbon film-supported copper grids. XPS core level emission spectra were acquired using a Surface Science Instrument SSX-100-301 spectrometer operating an Al K\textsubscript{$\alpha$} source ($h\nu$ = 1486.5 eV), achieving an overall energy resolution of 0.9 eV. The collected spectra were fitted using the XPSPEAK software, subtracting a Shirley-type background\cite{Shirley72} and fitting the peaks using mixed Gaussian and Lorentzian components. The samples were prepared by repeated dropcasting onto clean highly doped silicon substrates (with native oxide), and dried under vacuum between several drops and at the end of the deposition process.

\subsection{Computational Details} All calculations were performed with the CP2K\cite{2014cp2k,Kuhne20} program at the DFT level of theory using the Perdew-Burke-Ernzerhof (PBE) exchange and correlation functional.\cite{PBE1996} Second-generation dispersion corrections (D2)\cite{Grimme06} were used to take into account a proper description of van der Waals interactions. Goedecker-Teter-Hutter (GTH) pseudopotentials,\cite{GTH1996} together with double-zeta quality basis sets (DZVP), were used. The Gaussian-and-Plane-Waves (GPW) method as implemented in CP2K\cite{2005quickstep} was used; the energy cutoff for the auxiliary plane-wave basis was set to 340 Ry cutoff. The wavefunction convergence criterion was set to $10^{-6}$ Hartree. Geometry optimization of the systems was performed by using Broyden-Fletcher-Goldfarb-Shanno (BFGS) algorithm by setting a root mean square (RMS) value of $10^{-4}$ Hartree/Bohr for the force and $10^{-4}$ Bohr for the geometry as convergence criteria. The systems were treated as periodic. Reduced graphene oxide was modeled starting from the hexagonal graphene supercell (a = 1.98 nm) consisting of 164 carbon atoms. The Restrained Electrostatic Potential (RESP) charges\cite{Bayly93} for periodic systems were evaluated by using REPEAT methods\cite{2009RESP_REPEAT} as implemented in CP2K.\cite{Golze15} After structural minimization, the functionalized system was subject to free molecular dynamics (MD). MD simulations were performed in the NVT ensemble (T = 300 K) employing a canonical-sampling-through-velocity-escalating (CSVR) thermostat\cite{Bussi07} with a time constant of 200~fs. A single integration time step of 0.4 fs was used. MD simulations were carried out for 12 ps. Vibrational spectra were obtained by calculating the Fourier transform of the atoms velocity auto-correlation function (VACF) taken from the last 10 ps of the simulated trajectory of the system.\cite{Thomas13}

%
%
%
%


\phantomsection
\section*{Acknowledgments} 

\addcontentsline{toc}{section}{Acknowledgments} 

The authors thank Fabio Beltram for his continuous support. The authors also thank Pasqualantonio Pingue, Francesco Bonaccorso and BeDimensional Srl for the supply of graphene powder. Andrea Camposeo is gratefully acknowledged for the supply of rGO. The authors thank Giovanni Signore and Cinzia Cepek for useful discussions. This research was funded by EU-H2020 FETPROACT LESGO (Agreement No. 952068), and by the Italian Ministry of University and Research (project MONSTRE-2D PRIN2017 KFMJ8E).

\section*{Supporting Information}
Supporting Information is available online or from the authors.


\phantomsection
\bibliographystyle{unsrt}
\bibliography{bib.bib}


\end{document}